\documentclass[11pt]{article}
\usepackage{amsmath,amssymb,graphicx,epic}

\newcommand{\be}{\begin{eqnarray}}
\newcommand{\ee}{\end{eqnarray}}
\newcommand{\nn}{\nonumber}

\def\a{{\alpha}}
\def\b{{\beta}}
\def\g{{\gamma}}
\def\d{{\delta}}
\def\e{{\epsilon}}

\def\s{{\sigma}}

\def\hi{{\hat{\imath}}}
\def\hj{{\hat{\jmath}}}
\def\ta{{\tilde\alpha}}
\def\tb{{\tilde\beta}}

\def\reals{{\mathbb{R}}}
\def\ints{{\mathbb{Z}}}

\def\cD{{\cal{D}}}
\def\cE{{\cal{E}}}
\def\cG{{\cal{G}}}
\def\cL{{\cal{L}}}
\def\cV{{\cal{V}}}
\def\cQ{{\cal{Q}}}
\def\cP{{\cal{P}}}

\def\lae{{\mathfrak{e}}}
\def\rs{{\mathfrak{gl}(9)\oplus\mathfrak{so}(2,1)}}
\def\pt{{{\partial}_t}}

\begin{document}

{\flushright AEI-2004-112\\hep-th/0411225\\[1cm]}

\begin{center}
{\LARGE {\bf  \boldmath ${\rm IIB}$ supergravity 
  and $E_{10}$\unboldmath}}\\[1cm]
{\bf Axel Kleinschmidt and Hermann Nicolai}\\
Max--Planck--Institut f\"ur Gravitationsphysik
(Albert--Einstein--Institut)\\
M\"uhlenberg 1, D-14476 Golm, Germany\\
{\tt axel.kleinschmidt,hermann.nicolai@aei.mpg.de}\\[17mm]
\end{center}

\begin{center}
{\bf Abstract}\\[5mm]
\begin{tabular}{p{12cm}}
We analyse the geodesic $E_{10}/K(E_{10})$ $\s$-model in a level
decomposition w.r.t. the $A_8\times A_1$ subalgebra of $E_{10}$, adapted 
to the bosonic sector of type IIB supergravity, whose $SL(2,\reals)$ 
symmetry is identified with the $A_1$ factor. The bosonic supergravity 
equations of motion, when restricted to zeroth and first order spatial 
gradients, are shown to match with the $\s$-model equations of motion up 
to level $\ell =4$. Remarkably, the self-duality of the five-form field 
strength is implied by $E_{10}$ and the matching.\\[1cm]
\end{tabular}
\end{center}

\begin{section}{Introduction}

The simple and essentially unique geodesic Lagrangian describing a null
world line in the infinite-dimensional coset manifold $E_{10}/K(E_{10})$
has been shown to reproduce the dynamics of the bosonic sector of
eleven-dimensional supergravity in the vicinity of a space-like
singularity \cite{Damour:2002cu,Damour:2004zy}. This result was
subsequently extended to {\em massive} ${\rm IIA}$ supergravity in
\cite{Kleinschmidt:2004dy}, where also parts of the fermionic sector
were treated for the first time. A main ingredient in the derivation
of these results was the level decomposition of $E_{10}$ w.r.t. the
$A_9$ and $D_9$ subalgebras of $E_{10}$, respectively. Here, we
extend these results to type IIB supergravity, and
demonstrate that this model as well can be incorporated into the
$E_{10}/K(E_{10})$  $\sigma$-model within the framework proposed
in~\cite{Damour:2002cu}, by making a level decomposition w.r.t. the
$A_8 \times A_1$ subalgebra of $E_{10}$. As we will explain, this
decomposition is precisely adapted to type IIB supergravity, in that
it gives rise to the field representation content of IIB supergravity,
where the $A_1$ factor becomes identified with the continuous $SL(2,\reals)$
symmetry of the IIB theory. Furthermore, the bosonic supergravity
equations of motion, when restricted to zeroth and first order spatial
gradients, match with the $\s$-model equations of motion up to and
including level $\ell =4$. Perhaps our most important new result
here is that the self-duality of the five-form field strength is
implied by the dynamical matching between the  $E_{10}/K(E_{10})$
$\sigma$-model and the supergravity equations of motion, and does
not require local supersymmetry or some other extraneous argument
for its explanation.

Related results had been obtained previously in the framework
of another, and conceptually different proposal, according to which it
is the `very extended' Kac--Moody algebra $E_{11}$ that underlies $D=11$
supergravity or a suitable extension thereof \cite{West}. This proposal
can likewise be extended to massive IIA, and to IIB supergravity
\cite{Schnakenburg:2001ya,Schnakenburg:2002xx}, and consistency with
a level decomposition of the adjoint representation of $E_{11}$ was
shown in \cite{Kleinschmidt:2003mf}. By contrast, the present
construction shows that the hyperbolic Kac--Moody algebra $E_{10}$ is
already `big enough' by itself to accommodate all the maximally
supersymmetric theories in $D=10$ and $D=11$. Unlike $E_{10}$,  $E_{11}$ 
does not allow for an action unless one introduces an unphysical 
extraneous coordinate \cite{Englert2003}. In the absence of an action 
principle, the self-duality restriction on the the five-form field 
strength, as well as the mutual duality between the three- and 
seven-form field strengths must be imposed as an extra requirement.

Combining the known results, we can summarize the correspondence 
between the maximally supersymmetric theories and the maximal rank 
regular subalgebras of $E_{10}$ as follows
\be
A_9 \subset E_{10} \quad &\Longleftrightarrow& \qquad
  \mbox{$D=11$ supergravity} \nn\\
D_9 \subset E_{10} \quad &\Longleftrightarrow& \qquad
  \mbox{massive IIA supergravity} \nn\\
A_8 \times A_1 \subset E_{10} \quad &\Longleftrightarrow& \qquad
  \mbox{IIB supergravity} \nn
\ee
The decompositions of $E_{10}$ w.r.t. its other rank 9 regular subalgebras
$A_{D-2} \times E_{11-D}$ (for $D=3,\ldots,9$) will similarly reproduce 
the representation content of maximal supergravities in $D$ space-time 
dimensions as the lowest level representations. The first factor
here is identified with the $SL(D-1)$ acting on the spatial vielbein,
while $E_{11-D}$ is the Cremmer--Julia hidden symmetry \cite{CremmerJulia}.
The only missing, but perhaps the most interesting, piece in this analysis 
is the decomposition w.r.t. the {\em affine} subalgebra $E_9$ obtained 
in the reduction to two dimensions.

In \cite{Kleinschmidt:2004dy}, we have shown that, with the exception
of the space-filling D9-brane, $E_{10}$ can accommodate all D-branes,
such that NSNS and RR fields are associated with even and odd levels,
respectively, in the $D_9$ decomposition of $E_{10}$.\footnote{The
  possible relevance of the D9-brane for this algebraic analysis
  was emphasized to us by S. Chaudhuri.}
These results are
confirmed by the present investigation, but with the important difference
that NSNS and RR fields occur in the same $SL(2,\reals)$ multiplets, and
hence transform into one another under the action of $SL(2,\reals)$.
The precise assignment of these fields to parts of the $E_{10}$
structure is one of the main results of the present work.

This article is structured as follows. First, we deduce the generators 
of $E_{10}$ and their relations as appropriate for the $A_8\times A_1$
decomposition up to $\ell=4$. From this we deduce the $\s$-model
dynamics which are then shown to be equivalent to the reduced
IIB supergravity equations if the fields are identified in the
right way.

\end{section}

\begin{section}{$\lae_{10}$ relations in $\rs$ form}

Our analysis is based on a decomposition of the hyperbolic Kac--Moody
algebra $\lae_{10}$ under its $\rs$ subalgebra, as indicated in
figure \ref{e10dynk}. In the table below we list the field content
on the first five levels. All representations occur with outer 
multiplicity one. For the decomposition technique we refer readers to
\cite{Damour:2002cu,West:2002jj,FisNic03,Kleinschmidt:2003mf}.\\
\begin{center}
\begin{tabular}{|c|c|c|l|}
\hline
$\ell$&$A_8\oplus A_1$ representation&Generator&Interpretation\\
\hline\hline
1&$[0,0,0,0,0,0,1,0]\otimes {\bf 2}$&$E^{a_1a_2}{}_\a$&F1/D1\\
2&$[0,0,0,0,1,0,0,0]\otimes {\bf 1}$&$E^{a_1\ldots a_4}$&D3\\
3&$[0,0,1,0,0,0,0,0]\otimes {\bf 2}$&$E^{a_1\ldots a_6}{}_\a$& NS5/D5\\
4&$[0,1,0,0,0,0,0,1]\otimes {\bf 1}$&$E^{a_1\ldots a_7|a_8}$&KK Monopole\\
&$[1,0,0,0,0,0,0,0]\otimes {\bf 3}$&${E^{a_1\ldots a_8}}_i$&NS7/D7/*\\
5&$[1, 0, 0, 0, 0, 0, 1, 0]\otimes {\bf 2}$&&?\\
&$[0, 0, 0, 0, 0, 0, 0, 1]\otimes {\bf 2}$&&?\\
\hline
\end{tabular}
\end{center}
The $\mathfrak{sl}(2,\reals)\cong\mathfrak{so}(2,1)$ representation is
labelled by its dimension rather than by its Dynkin labels for ease of
notation. For $\ell=0$ the content is simply the adjoint representation
of the subalgebra $\rs$ (which includes the Cartan subalgebra generator 
$h_0$ of the omitted black node). We have also listed the generators 
for the first four levels, whose complete commutation relations will be 
worked out below.

In the last column of the table we indicate the interpretation of these
representations in terms of branes of the ${\rm IIB}$ theory, as suggested
by the representation structure and the study of finite-dimensional
U-duality groups \cite{Obers:1998fb}. Level $\ell=0$ contains the 
$SL(2,\reals)/SO(2)$ coset degrees of freedom which in supergravity are 
parametrized by the axion (the source of the D($-1$) instanton) and the 
dilaton, and a third $SO(2)$ gauge degree of freedom which can be gauged
to zero. At level $\ell=1$ we find the fundamental (F1) string and Dirichlet 
(D1) string, which transform as a doublet under $SL(2,\reals)$ 
\cite{Schwarz:1995dk}. Evidently the $SL(2,\reals)$ symmetry thus mixes 
NSNS and RR type fields. This feature persists for the other
D$p$-branes of ${\rm IIB}$ string theory which couple to the $(p+1)$-form 
RR potentials, with the exception of the rank-4 potential on $\ell=2$
which is an $SL(2,\reals)$ singlet, consistent with the S-self-duality
of the D$3$-brane. For $\ell=3$ we get a doublet of five-branes, while at 
$\ell=4$ we find a triplet of seven-branes, the best known of which
is the D7-brane (other members of the multiplet have also been investigated
in the literature \cite{Meessen:1998qm,Bergshoeff:2002mb}). 
The asterisk indicates a sublety for the seven branes arising from issues of 
gauge fixing: the corresponding fields are here interpreted as dual to the 
spatial gradients of the three $\ell=0$ scalars, hence the gauge
fixing implies a constraint on the charges associated with the three
generators. We have also included the level $\ell=5$ 
representations because this is the lowest level where a difference 
between $E_{10}$ and $E_{11}$ occurs, and because of their relevance
to nine-branes. There the situation is much less clear, and will
be commented upon in the concluding section.

\begin{figure}
\begin{center}
\scalebox{1}{
\begin{picture}(340,60)
\put(5,-5){$\alpha_1$}
\put(45,-5){$\alpha_2$}
\put(85,-5){$\alpha_3$}
\put(125,-5){$\alpha_4$}
\put(165,-5){$\alpha_5$}
\put(205,-5){$\alpha_6$}
\put(245,-5){$\alpha_7$}
\put(285,-5){$\alpha_0$}
\put(325,-5){$\alpha_9$}
\put(260,45){$\alpha_{8}$}
\thicklines
\multiput(10,10)(40,0){9}{\circle{10}}
\multiput(15,10)(40,0){8}{\line(1,0){30}}
\put(250,50){\circle{10}} \put(250,15){\line(0,1){30}}
\put(290,10){\circle*{10}}
\end{picture}}
\caption{\label{e10dynk}\sl Dynkin diagram of $\lae_{10}$ with nodes
  marked for a decomposition under $\rs$.}
\end{center}
\end{figure}

We now study the commutation relations of the corresponding
generators up to $|\ell|=4$.\footnote{The commutators for the positive level
  ($0\leq\ell\leq 3$) generators and two of the three anti-symmetric
  eight-forms on $\ell=4$ were already given in \cite{Schnakenburg:2001ya}, 
  with the generators in representations of $GL(10)$, but without manifest 
  $SL(2,\reals)$ covariance. The algebra $G_{{\rm IIB}}$ introduced there 
  is a truncation of the $\ell\geq 0$ Borel subalgebra of $E_{11}$, but 
  does not correspond to a consistent truncation of $E_{11}$ if the 
  negative level generators are also included.}

{\bf Level $\ell=0$:} The generators are $K^a{}_b$ and $J_i$ with
relations
\be
[K^a{}_b,K^c{}_d] &=& \d^c_b K^a{}_d - \d^a_d K^c{}_b,\\{}
[J_i,J_j] &=& \e_{ij}{}^k J_k,\\{}
[J_i,K^a{}_b] &=&0.
\ee
The indices take values $a=1,\ldots,9$; $i=1,2,3$ and the
$\mathfrak{so}(2,1)$ metric is $\eta=\textrm{diag}(-,+,-)$ such that
explicitly
\be
[J_1,J_2]=-J_3,\qquad [J_2,J_3]=-J_1,\qquad [J_3,J_1]=J_2.
\ee
The identification with the Chevalley basis of $\lae_{10}$ on the
subalgebra nodes is
\be\label{slid}
e_a = K^a{}_{a+1},\qquad f_a = K^{a+1}{}_a,\qquad h_a = K^a{}_a -
K^{a+1}{}_{a+1},
\ee
for $a=1,\ldots,8$. The trace $K\equiv\sum_{a=1}^9 K^a{}_a$ is
\be\label{tr}
K=-8h_1-16h_2-24h_3-32h_4-40h_5-48h_6-56h_7-28h_8-18h_9-36h_0.
\ee
For the `decoupled' $\mathfrak{sl}(2,\reals)$ node 9, the identifications are
\be\label{id9}
e_9=J_+,\qquad f_9=J_-,\qquad h_9=2J_3,
\ee
where $J_\pm= J_1\pm J_2$. The maximal compact subalgebra $\mathfrak{ke}_{10}
\subset\lae_{10}$ consists of all `antisymmetric' elements of $\lae_{10}$,
where the generalized transposition is defined as
\be
x^T := - \theta (x) \quad , \qquad x\in \lae_{10},
\ee
and $\theta$ is the Chevalley involution. With this definition, the
antisymmetric elements at level $\ell =0$ are ${K^a}_b - {K^b}_a$ and
$J_2$; they generate the compact level $\ell=0$ subalgebra
$\mathfrak{so}(9)\oplus\mathfrak{so}(2)\subset\mathfrak{ke}_{10}$.
The symmetric elements at $\ell=0$ are ${K^a}_b + {K^b}_a$, and $J_1$
and $J_3$.

A two-dimensional representation of $\mathfrak{so}(2,1)$
is furnished by the Pauli matrices:
\be
\s_1=\left(\begin{array}{cc}0&1\\1&0\end{array}\right),\qquad
\s_2=\left(\begin{array}{cc}0&1\\-1&0\end{array}\right),\qquad
\s_3=\left(\begin{array}{cc}1&0\\0&-1\end{array}\right).\qquad
\ee
We introduce indices $\a,\b=1,2$ for this representation by writing
$(\s_i)^\a{}_\b$. The tensor $\e^{\a\b}$ equals $\s_2$ as a set of
numbers and satisfies the condition $\e^{\a\g}\e_{\b\d}(\s_i)^\d{}_\g
=(\s_i)^\a{}_\b$. The inverse $\e_{\a\b}=-\e^{\a\b}$ satisfies
$\e^{\a\b}\e_{\b\gamma}=\d^\a_\gamma$. We also define
$\s^i\equiv\eta^{ij}\s_j$.

The normalizations of the generators within $\lae_{10}$ are
\be
\langle K^a{}_b|K^c{}_d\rangle = \d^a_d\d^c_b-\d^a_b\d^c_d,\qquad
\langle J_i|J_j\rangle = -\frac12 \eta_{ij}.
\ee
consistent with the standard normalization of the Chevalley generators.
From this it can be shown that all compact (i.e. anti-symmetric)
generators have negative norm as expected.

{\bf Level $\ell=1$:} The only representation is an
$\mathfrak{so}(2,1)$ doublet of $\mathfrak{gl}(9)$ two-forms, denoted
by $E^{ab}{}_\a$ and transforming under $\rs$ as
\be
[K^a{}_b,E^{cd}{}_\a] = -2\d^{[c}_b E^{d]a}{}_\a,\qquad
[J_i,E^{cd}{}_\a] = \frac12 (\s_i)^\b{}_\a E^{cd}{}_\b.
\ee
The transposed field is
\be
F_{ab}{}^\a := (E^{ab}{}_\a)^T
\ee
and satisfies
\be
[K^a{}_b,F_{cd}{}^\a] = 2\d^a_{[c} F_{d]b}{}^\a,\qquad
[J_i,F_{cd}{}^\a] = -\frac12 (\s_i)^\a{}_\b F_{cd}{}^\b.
\ee
The identification of the remaining Chevalley generators yields
\be
e_0 = E^{89}{}_2 \quad,\quad f_0= F_{89}{}^2 \quad,\quad
h_0= -\frac14 K +K^8{}_8 +K^9{}_9- J_3,
\ee
where $h_0$ is already identified in $\rs$ through
(\ref{slid})--(\ref{id9}).

Demanding the normalization
\be
\langle E^{ab}{}_\a|F_{cd}{}^\b\rangle = 2 \d^{ab}_{cd}\d^\b_\a
\ee
leads to the commutator
\be
[E^{ab}{}_\a,F_{cd}{}^\b] = -\frac12 \d^\b_\a \d^{ab}_{cd} K +
4\d^\b_\a \d^{[a}_{[c} K^{b]}{}_{d]} -2 \d^{ab}_{cd}(\s^i)^\b{}_\a
J_i.
\ee

{\bf Level $\ell=2$:} The only representation is an $\mathfrak{so}(2,1)$
singlet, transforming as an antisymmetric rank four tensor under
$\mathfrak{gl}(9)$. We denote it by $E^{a_1\ldots a_4}$, and its
transpose by $F_{a_1\ldots a_4}:=(E^{a_1\ldots a_4})^T$. They are
obtained by commuting two $|\ell|=1$ elements:
\be
[E^{ab}{}_\a,E^{cd}{}_\b] =  \e_{\a\b} E^{abcd},\qquad
[F_{ab}{}^\a,F_{cd}{}^\b] =  \e^{\a\b} F_{abcd}.
\ee
The normalization of commutators is consistent with
\be
\langle E^{a_1\ldots a_4}|F_{b_1\ldots b_4}\rangle = 4! \, \d^{a_1\ldots
  a_4}_{b_1\ldots b_4}.
\ee
The remaining commutation relations up to $|\ell|\leq 2$ are
\be
[E^{a_1\ldots a_4},F_{b_1b_2}{}^\b] &=& 12
\e^{\b\a}\d^{[a_1a_2}_{b_1b_2} E^{a_3a_4]}{}_\a,\\{}
[F_{b_1\ldots b_4},E^{a_1a_2}{}_\a] &=& 12
\e_{\a\b}\d^{a_1a_2}_{[b_1b_2}F_{b_3b_4]}{}^\b,\\{}
[E^{a_1\ldots a_4},F_{b_1\ldots b_4}] &=& -12
\d^{a_1\ldots a_4}_{b_1\ldots b_4} K + 96 \d^{[a_1\ldots
    a_3}_{[b_1\ldots b_3} K^{a_4]}{}_{b_4]}.
\ee
Note that there is no term proportional to $J_i$ in the last commutator
which in is agreement with the ${\rm IIB}$ supergravity structure as we
will see.

{\bf Level $\ell=3$:} The only representation is an $\mathfrak{so}(2,1)$
doublet of six-forms under $\mathfrak{gl}(9)$. The generators
$E^{a_1\ldots a_6}{}_\a$ and their transpose $F_{a_1\ldots a_6}{}^\a$
are obtained via
\be
[E^{a_1a_2}{}_\a, E^{a_3\ldots a_6}] = E^{a_1\ldots
  a_6}{}_\a,\qquad
[F_{a_1a_2}{}^\a, F_{a_3\ldots a_6}] = - F_{a_1\ldots
  a_6}{}^\a,
\ee
consistent with the normalization
\be
\langle E^{a_1\ldots a_6}{}_\a| F_{b_1\ldots b_6}{}^\b\rangle =
6! \, \d^\b_\a \d^{a_1\ldots a_6}_{b_1\ldots b_6}.
\ee
The remaining relations are
\be
[E^{a_1a_2}{}_\a, F_{b_1\ldots b_6}{}^\b] &=& -30
  \d^\b_\a \d^{a_1a_2}_{[b_1b_2} F_{b_3\ldots b_6]},\\{}
[F_{b_1b_2}{}^\b, E^{a_1\ldots a_6}{}_\a] &=& 30
  \d^\b_\a \d^{[a_1a_2}_{b_1b_2} E^{a_3\ldots a_6]},\\{}
[E^{a_1\ldots a_4}, F_{b_1\ldots b_6}{}^\b] &=& 360
  \d^{a_1\ldots a_4}_{[b_1\ldots b_4}F_{b_5b_6]}{}^\b,\\{}
[F_{b_1\ldots b_4}, E^{a_1\ldots a_6}{}_\a] &=& -360
  \d^{[a_1\ldots a_4}_{b_1\ldots b_4}E^{a_5a_6]}{}_\a,\\{}
[E^{a_1\ldots a_6}{}_\a, F_{b_1\ldots b_6}{}^\b] &=&
-540\d^\b_\a \d^{a_1\ldots a_6}_{b_1\ldots b_6} K +
4320 \d^\b_\a \d^{[a_1\ldots a_5}_{[b_1\ldots b_5}
    K^{a_6]}{}_{b_6]}\nn\\&&\quad\quad\quad\quad\quad\quad
 - 720 \d^{a_1\ldots a_6}_{b_1\ldots b_6}
   (\s^i)^\b{}_\a J_i.
\ee

{\bf Level $\ell=4$:} The fields are a tensor of mixed $A_8$ Young symmetry,
usually called the `dual graviton', transforming as a singlet under
$\mathfrak{so}(2,1)$, and a fully anti-symmetric $A_8$ eight-form,
transforming as a triplet under $\mathfrak{so}(2,1)$. They are
obtained by
\be
[E^{a_1a_2}{}_\a,E^{a_3\ldots a_8}{}_\b] = \frac16
\e_{\a\b}E^{a_1a_2[a_3\ldots a_7|a_8]} +  E^{a_1\ldots a_8}{}_i
(\e\s^i)_{\a\b} .
\ee
Therefore
\be
E^{a_1\ldots a_8}{}_i &=&-\frac12 (\s_i\e)^{\a\b}
\big[E^{[a_1a_2}{}_\a,E^{a_3\ldots a_8]}{}_\b\big],\\
E^{a_1\ldots a_7|a_8} &=& -63 \, \e^{\a\b}
\big[E^{[a_1a_2}{}_\a,E^{a_3\ldots a_7]a_8}{}_\b\big].
\ee
This is consistent with the normalizations
\be
\langle E^{a_1\ldots a_8}{}_i|F_{b_1\ldots b_8}{}^j\rangle &=& \frac12\cdot 8!
\, \d^j_i \d^{a_1\ldots a_8}_{b_1\ldots b_8},\\
\langle E^{a_1\ldots a_7|a_8}|F_{b_1\ldots b_7|b_8}\rangle &=&
\frac{7\cdot 7!}{8}\,\big(\d^{a_1\ldots a_7}_{b_1\ldots b_7}
\d^{a_8}_{b_8} + \d^{[a_1}_{b_8} \d^{a_2\ldots a_7]}_{[b_1\ldots b_6}
  \d^{a_8}_{b_7]}\big).
\ee
The remaining commutation relations for the second representation
$E^{a_1\ldots a_8}{}_i$ are
\be
\big[ E^{a_1\ldots a_8}{}_i, F_{b_1b_2}{}^\b \big] &=& 28
(\s_i\e)^{\b\g} \d^{[a_1a_2}_{b_1b_2} E^{a_3\ldots
    a_8]}{}_\g,  \\
\big[ E^{a_1\ldots a_8}{}_i, F_{b_1\ldots b_4}\big] &=& 0,  \\
\big[ E^{a_1\ldots a_8}{}_i, F_{b_1\ldots b_6}{}^\b\big] &=& 
-\frac14\cdot 8!\, (\s_i\e)^{\b\g} \d^{[a_1\ldots a_6}_{b_1\ldots b_6}
  E^{a_2a_8]}{}_\g,\\
\big[ E^{a_1\ldots a_8}{}_i, F_{b_1\ldots b_8}{}^j] &=& -\frac12\cdot
8! \d^j_i \d^{a_1\ldots a_8}_{b_1\ldots b_8} K +4 \cdot 8! \d^j_i
  \d^{[a_1\ldots a_7}_{[b_1\ldots b_7} K^{a_8]}{}_{b_8]} -\frac12
  \cdot 8! \, \e_i{}^{jk} J_k,
\ee
and the corresponding relations for the transposed fields. We note
that the anti-symmetrized commutator $[E^{[a_1\ldots a_4},E^{a_5\ldots
      a_8]}]$ vanishes, consistent with the $E_7^{++}$ subsector.

The commutation relations involving $E^{a_1\ldots a_7|a_8}$ are
\be
\big[E^{a_1\ldots a_7|a_8}, F_{b_1b_2}{}^\b\big] &=& +378
  \e^{\b\a} (\d_{b_1 b_2}^{[a_1a_2} E^{a_3\ldots a_7]a_8}{}_\a
  +\d_{b_1b_2}^{a_8[a_1}E^{a_2\ldots a_7]}{}_\a),\\
\big[E^{a_1\ldots a_7|a_8}, F_{b_1\ldots b_4}\big] &=& 1890
  (\d_{b_1\ldots b_4}^{[a_1\ldots a_4}E^{a_5\ldots a_7]a_8}
  +\d_{b_1\ldots b_4}^{a_8[a_1\ldots a_3}E^{a_4\ldots a_7]}),\\
\big[E^{a_1\ldots a_7|a_8}, F_{b_1\ldots b_6}{}^\b\big] &=&
45360 \e^{\b\a} (\d_{b_1\ldots b_6}^{[a_1\ldots a_6} E^{a_7]a_8}{}_\a
  +\d_{b_1\ldots b_6}^{a_8[a_1\ldots a_5}E^{a_6a_7]}{}_\a),\\
\big[X_{a_1\ldots a_7|a_8}E^{a_1\ldots a_7|a_8}, F_{b_1\ldots
    b_7|b_8}\big] &=& -7!\big( X_{b_1\ldots b_7|b_8} K -X_{b_1\ldots
  b_7|c} K^c{}_{b_8}\nn\\
&&\quad\quad\quad - 7 K^c{}_{[b_1} X_{b_2\ldots b_7]c|b_8}^{\ }\big),
\ee
where we have introduced an auxiliary tensor $X_{a_1\ldots a_7|a_8}$ 
to simplify the expression in the last line on the right hand
side. Besides the transposed relations of the above, we also find
\be
[E^{a_1\ldots a_8}{}_i,F_{b_1\ldots b_7|b_8}] =0.
\ee

\end{section}

\begin{section}{$\s$-model equations of motion}

In this section, we work out the $\s$-model equations of motion, using
the formulation in terms of $K(E_{10})$ orthonormal frames developed
in \cite{Damour:2004zy} and \cite{Kleinschmidt:2004dy} for the $A_9$
and $D_9$ level decompositions of $E_{10}$, respectively.
Accordingly, we parametrize the coset space in terms of a `matrix'
$\cV \equiv \cV\big(A(t)\big) \in E_{10}$. Here $A(t)$ are `local
coordinates' on the infinite dimensional coset manifold $E_{10}/K(E_{10})$.
Making use of the local $K(E_{10})$ invariance, a convenient choice of
gauge is the Borel type triangular gauge, with the fields $A=A^{(\ell)}$
for $\ell\geq 0$ to parametrize the $E_{10}/K(E_{10})$ coset space.
The scalar fields $A(t)$ couple via the Lie algebra-valued `velocity'
\be
\pt\cV \cV^{-1} &=& P^{(0)}_{ab} S^{ab} + Q^{(0)}_{ab} J^{ab} +
  Q^{(0)} J_2 +  P^{(0)\hi} J_\hi
+\frac12 P_{a_1a_2}^{(1)}{}^\a E^{a_1a_2}{}_\a\nn\\
&& + \frac1{4!} P_{a_1\ldots
  a_4}^{(2)}E^{a_1\ldots a_4}
 +\frac1{6!} P_{a_1\ldots a_6}^{(3)}{}^\a E^{a_1\ldots a_6}{}_\a +
  \frac1{7!} P_{a_1\ldots a_7|a_8}^{(4)}E^{a_1\ldots a_7|a_8}\nn\\&& +
 \frac1{8!} P_{a_1\ldots a_8}^{(4)}{}^i E^{a_1\ldots a_8}{}_i+\ldots
 \quad \in \lae_{10}
\ee
where hatted indices $\hat{\imath}$ label the $SL(2,\reals)/SO(2)$ coset
generators and hence take only the values ${\hat{\imath}}=1,3$:
\be
P^{(0)\hi} J_\hi \equiv  P^{(0)1} J_1 + P^{(0)3} J_3.
\ee
Splitting the `velocity' as $\pt\cV \cV^{-1}=\cQ+\cP$, where
$\cQ\in\mathfrak{ke}_{10}$ is the $K(E_{10})$ gauge connection
and $\cP \in \lae \ominus \mathfrak{ke}_{10}$ in the coset, we write
\be
\cQ &=& Q^{(0)}_{ab} J^{ab} + Q^{(0)} J_2+
 \sum_{\ell>0} P^{(\ell)}\star \frac12(E^{(\ell)}-F^{(\ell)}),\\{}
\cP &=& P^{(0)}_{ab} S^{ab} + P^{(0)\hi} J_\hi
 +\sum_{\ell>0} P^{(\ell)}\star \frac12(E^{(\ell)}+F^{(\ell)}),
\ee
with $F^{(\ell)}:=(E^{(\ell)})^T$, $J^{ab}=\frac12(K^a{}_b-K^b{}_a)$ and
$S^{ab}=\frac12(K^a{}_b+K^b{}_a)$ and the higher level contributions
  are indicated schematically.

Following \cite{Damour:2004zy} we define the `covariant' derivative
for $\ell>0$
\be
&&\cD^{(0)}P^{(\ell)}\star \frac12(E^{(\ell)}+F^{(\ell)})\nn\\
&&\quad=  \pt P^{(\ell)}\star\frac12(E^{(\ell)}+F^{(\ell)})
-\big[Q^{(0)}_{ab}J^{ab},P^{(\ell)}\star\frac12(E^{(\ell)}+F^{(\ell)})\big]
 \nn\\
&&\quad\quad-\big[P^{(\ell)}\star\frac12(E^{(\ell)}-F^{(\ell)}),
P^{(0)}_{ab}S^{ab}\big]\\
&&\quad\quad -\big[Q^{(0)} J_2,P^{(\ell)}\star\frac12(E^{(\ell)}
+ F^{(\ell)})\big]
 -\big[P^{(\ell)}\star\frac12(E^{(\ell)}-F^{(\ell)}),P^{(0)\hi} J_\hi\big].\nn
\ee
This expression is covariant with respect to both $\mathfrak{so}(9)$ and
$\mathfrak{so}(2)$. The analogous covariant derivatives for $\ell=0$
are
\be
(\cD^{(0)}P^{(0)}_{ab})S^{ab} &=& \pt P^{(0)}_{ab} S^{ab}
  - [Q^{(0)}_{ab}J^{ab},P^{(0)}_{cd}S^{cd}],\\
(\cD^{(0)}P^{(0)\hi}) J_\hi &=& \pt P^{(0)\hi} J_\hi
  - [Q^{(0)}J_2,P^{(0)\hi} J_\hi].
\ee
This reflects the fact that there are no terms coupling the orthogonal
summands $\mathfrak{so}(9)$ and $\mathfrak{so}(2)$.

The geodesic $\s$-model Lagrangian reads
\be
\cL(t)=\frac14 {n(t)}^{-1} \langle \cP(t)|\cP(t)\rangle
\ee
where $t$ is an affine parameter (`time'), and $\langle . | .\rangle$ 
is the standard bilinear form on the Kac Moody algebra. This Lagrangian
is unique because for infinite dimensional Kac Moody algebras the only
invariant form is quadratic. The above Lagrangian gives rise to the 
equation of motion
\be
n \pt(n^{-1} \cP) = [\cQ,\cP].
\ee
The Lagrange multiplier (`lapse') $n$ is needed for invariance under
reparametrizations of the time coordinate $t$ and ensures that 
the motion on the coset manifold takes place on a null geodesic.
In the truncation to $|\ell|\leq4$ we set
\be\label{trunc}
P^{(5)} = P^{(6)} = \dots = 0
\ee
With the commutation relations derived above we find
\be
\label{sigmaeomgrav}
n\cD^{(0)}(n^{-1}P^{(0)}_{ab}) &=& -\frac1{16}\d_{ab}
  P_{cd}^{(1)}{}^\a P_{cd}^{(1)}{}^\a
  + \frac12 P_{ac}^{(1)}{}^\a P_{bc}^{(1)}{}^\a
  - \frac1{96}\d_{ab} P_{c_1\ldots c_4}^{(2)} P_{c_1\ldots c_4}^{(2)}\nn\\
&&\quad +\frac1{12} P_{ac_1\ldots c_3}^{(2)} P_{bc_1\ldots
  c_3}^{(2)} -\frac1{16\cdot5!}\d_{ab}
  P_{c_1\ldots c_6}^{(3)}{}^\a  P_{c_1\ldots  c_6}^{(3)}{}^\a\nn\\
&&\quad + \frac1{2\cdot 5!} P_{ac_1\ldots c_5}^{(3)}{}^\a
  P_{bc_1\ldots c_5}^{(3)}{}^\a-\frac1{2\cdot 7!}\d_{ab} P_{c_1\ldots
  c_7|c_8}^{(4)} P_{c_1\ldots c_7|c_8}^{(4)}\nn\\
&&\quad+ \frac1{2\cdot 7!} P_{c_1\ldots c_7|a}^{(4)} P_{c_1\ldots
  c_7|b}^{(4)} + \frac1{2\cdot 6!} P_{ac_1\ldots c_6|d}^{(4)} P_{bc_1\ldots
  c_6|d}^{(4)}\nn\\
&&\quad - \frac1{4\cdot 8!}\d_{ab} P_{c_1\ldots c_8}^{(4)}{}^i
  P_{c_1\ldots c_8}^{(4)}{}^i + \frac1{4\cdot 7!} P_{ac_1\ldots
  c_7}^{(4)}{}^i   P_{bc_1\ldots c_7}^{(4)}{}^i ,\\
\label{sigmaeomscalars}
n\cD^{(0)}(n^{-1}P^{(0)\hi}) &=& \left(-\frac14
  P_{ab}^{(1)}{}^\a P_{ab}^{(1)}{}^\b
  -\frac1{2\cdot 6!} P_{a_1\ldots a_6}^{(3)}{}^\a
  P_{a_1\ldots a_6}^{(3)}{}^\b\right) (\s^{\hat{\imath}})^\b{}_\a\nn\\
&&\quad +\frac1{2\cdot 8!}\e^{\hi2\hj} P_{c_1\ldots c_8}^{(4)}{}^2\,
  P_{c_1\ldots  c_8}^{(4)}{}^\hj  \\
\label{sigmaeoml1}
n\cD^{(0)}(n^{-1}P_{ab}^{(1)}{}^\a) &=&
  \frac14 \e_{\a\b} P_{abcd}^{(2)}P_{cd}^{(1)}{}^\b +\frac1{2\cdot 4!}
  P_{abc_1\ldots c_4}^{(3)}{}^\a P_{c_1\ldots c_4}^{(2)}\nn\\
&&\quad+\frac1{160}
  \e^{\a\b} \left(P^{(3)}_{c_1\ldots c_6}{}^\b P^{(4)}_{c_1\ldots c_6[a|b]}
  + P^{(3)}_{c_1\ldots c_6}{}^\b P^{(4)}_{a[bc_1\ldots c_5|c_6]}\right)\nn\\
&&\quad - \frac{1}{4\cdot 6!} (\s_i\e)^{\a\b} P^{(4)}_{c_1\ldots c_6 ab}{}^i
  P^{(3)}_{c_1\ldots c_6}{}^\b,\\
  \label{sigmaeoml2}
n\cD^{(0)}(n^{-1}P_{a_1\ldots a_4}^{(2)}) &=& -\frac14 P_{a_1\ldots a_4
  b_1b_2}^{(3)}{}^\a P_{b_1b_2}^{(1)}{}^\b \e_{\a\b}\nn\\
&&\quad +\frac1{128}\left(P^{(2)}_{c_1\ldots c_4} P^{(4)}_{c_1\ldots c_4
  [a_1\ldots a_3|a_4]} +P^{(2)}_{c_1\ldots c_4} P^{(4)}_{a_1\ldots a_4
  c_1\ldots c_3|c_4}\right),\\
\label{sigmaeoml3}
n\cD^{(0)}(n^{-1}P_{a_1\ldots a_6}^{(3)}{}^\a) &=&
  -\frac3{160}\e^{\a\b} \left(P^{(2)}_{c_1c_2}{}^\b
  P^{(4)}_{c_1c_2[a_1\ldots a_5|a_6]} + P^{(2)}_{c_1c_2}{}^\b
  P^{(4)}_{c_1 a_1\ldots a_6|c_2}\right)\nn\\
&&\quad +\frac{1}{8}(\e\s_i)^{\a\b} P_{a_1\ldots a_6
    c_1c_2}^{(4)}{}^i P_{c_1c_2}^{(1)}{}^\b,\\
\label{sigmaeoml4a}
n\cD^{(0)}(n^{-1}P_{a_1\ldots a_7|a_8}^{(4)}) &=& 0,\\
\label{sigmaeoml4b}
n\cD^{(0)}(n^{-1}P_{a_1\ldots a_8}^{(4)}{}^i) &=& 0.
\ee
Indices on the same level are contracted with the Euclidean flat
metrics of $\mathfrak{so}(9)$ and $\mathfrak{so}(2)$; in particular,
the indices $\a,\b,\dots$ are no longer contracted in an $SL(2)$
invariant way.

The consistency of the truncation (\ref{trunc}) is ensured by the same 
arguments as in \cite{Damour:2004zy}. Note that although this
requires only a finite number of non-vanishing $P^{(\ell)}$, the
`unendlichbein' $\cV$ parametrized by the $E_{10}/K(E_{10})$ coset
coordinates $A(t)$ needs to evolve correctly to ensure the vanishing
of $P^{(\ell)}$ for $|\ell|>4$. This involves the full structure of
$E_{10}$.

\end{section}

\begin{section}{Comparison with ${\rm IIB}$ supergravity}

We now compare the equations (\ref{sigmaeomgrav})--(\ref{sigmaeoml4b})
with the type ${\rm IIB}$ supergravity equations of motion
\cite{Schwarz:1983wa,Schwarz:1983qr}. We will use conventions similar
to those of \cite{Bergshoeff:1995as,Cremmer:1997xj} in order to make
the $SL(2,\reals)$ invariance transparent.

As is well known, IIB supergravity requires the following bosonic
fields. For the zehnbein we choose a pseudo-Gaussian gauge with
lapse $N$ and vanishing shift
\be
{E_M}^A = \left(\begin{array}{cc}N &0 \\0& {e_m}^a \end{array}\right).
\ee
In addition, there are an $SL(2,\reals)$ doublet of two-form potentials
$A_{2,\ta}$ ($\ta=1,2$) and an $SL(2,\reals)$ singlet four-form $B_4$ with
self-dual field strength given by
\be\label{h5def}
H_5=dB_4 +\frac14\e^{\ta\tb}A_{2,\ta}dA_{2,\tb} =\, ^* \! H_5
\ee
We will drop the rank indices on $H_5$ and $dA_2$ in the remainder.
Last but not least, there are two scalar fields $\phi$ and $\chi$ (dilaton
and axion) which, in a convenient triangular gauge, parametrize the coset
$SL(2,\reals)/SO(2)$ according to  \cite{Cremmer:1997xj}
\be
\cE = \left(\begin{array}{cc}e^{\phi/2}&\chi e^{\phi/2}\\0&e^{-\phi/2}
\end{array}\right).
\ee
The matrix $\cE_\a{}^\ta$ can be thought of as an internal zweibein, which
serves to convert global $SL(2,\reals$) indices $\ta,\tb,\dots$ into 
local $SO(2)$ indices $\a,\b,\dots$, in complete analogy with the 
spatial neunbein ${e_m}^a$ which converts global $GL(9)$ indices to 
local (Lorentz) $SO(9)$ indices, and vice versa. Thus, the model possesses
a local $\mathfrak{so}(9)\oplus\mathfrak{so}(2)$ symmetry, which will be
directly identified with the $\mathfrak{so}(9)\oplus\mathfrak{so}(2)$
subalgebra of $\lae_{10}$ in the $A_8\times A_1$ decomposition of $E_{10}$
developed in the foregoing section. The scalar field $\phi$ and $\chi$
appear in the IIB supergravity Lagrangian via
\be
\partial_M \cE \cE^{-1} = R_M J_2 + S^\hi_M J_\hi
= \partial_M \phi J_2 + \partial_M \chi e^{\phi} J_+
\ee
The derivation of the IIB equations of motion is greatly facilitated
by the $SL(2,\reals)$ symmetry which fixes many couplings
uniquely. Converting the global indices to local indices we write for
the doublet of two-forms
\be
(\cE dA)_{ABC\,\a} = \cE_\a{}^\ta E_A{}^M E_B{}^N E_C{}^P
\partial_{[M} A_{NP],\ta},
\ee
such that
\be
(\cE dA)_{ABC\, 1} &=& e^{\phi/2} (dA)_{ABC,\tilde{1}}+\chi
e^{\phi/2} (dA)_{ABC,\tilde{2}},\\
(\cE dA)_{ABC\, 2} &=& e^{-\phi/2} (dA)_{ABC,\tilde{2}},
\ee
where $\tilde{1}$ and $\tilde{2}$ are global $SL(2,\reals)$ indices.

For the comparison with the above $\s$-model equations of motion
it is most convenient to write the bosonic field equations of IIB
supergravity in terms of {\em flat} $SO(1,9)\times SO(2)$ indices, viz.
\be
\label{iibgrav}
R_{AB} &=& -\frac14 S^\hi_A S_{\hi B} +\frac1{48}H_A{}^{C_1\ldots
  C_4}H_{BC_1\ldots C_4} -\frac1{480} \eta_{AB} H^{C_1\ldots
  C_5}H_{C_1\ldots C_5}\nn\\
&& +\frac14 (\cE dA)_{A}{}^{CD\a} (\cE dA)_{BCD}{}^\a
  -\frac1{48}\eta_{AB} (\cE dA)^{BCD\a} (\cE dA)_{BCD}{}^\a,\\
\label{iibcos}
D^A S_A^\hi &=& \frac1{12} (\cE dA)_{BCD}{}^\a (\cE
dA)^{BCD\b}(\s^\hi)^\a{}_\b,
\ee
for the Einstein equation and the coset $SL(2,\reals)/SO(2)$. $D^A$ 
is the $SO(1,9)\times SO(2)$ covariant derivative; splitting the 
divergence into time and space components, it reads
\be\label{covder}
D^A S_A^\hi &=& \partial^0  S_0^\hi+ \partial^a  S_a^\hi +
\omega^0{}_0{}^a S_a^\hi + \omega^a{}_a{}^0 S_0^\hi +\omega^a{}_a{}^b
S_b^\hi\nn\\ 
&&\quad+\frac12
  \e^{\hi 2\hj} (R^0 S_0^\hj + R^a S_a^\hj ),
\ee
where $\omega_{A\ BC}$ is the usual spin connection. For the form potentials
we obtain 
\be
\label{iibfive}
D^A H_{AC_1\ldots C_4} &=& -\frac1{144} \e_{\a\b} \e_{C_1\ldots
  C_4}{}^{D_1\ldots D_6} (\cE dA)_{D_1D_2D_3}{}^\a (\cE
  dA)_{D_4D_5D_6}{}^\b,\\
\label{iibtwo}
D^A(\cE dA)_{ABC}{}^\a &=& -\frac14 \e^{\hi2\hj} S^{A\hi}
  (\s_\hj)^\a{}_\b (\cE dA)_{ABC}{}^\b
  +\frac1{12} H_{BC}{}^{DEF} \e^{\a\b}  (\cE dA)_{DEF}{}^\b.
\ee
Here, it is not necessary to restrict
$H_{C_1\ldots C_5}$ to be self-dual. This condition can be
imposed additionally by hand since it is consistent with the Bianchi
identity for $B_4$ from (\ref{h5def}):
\be
D_{[A_1}H_{A_2\ldots A_6]} = -\frac14\e_{\a\b}(\cE dA)_{[A_1A_2A_3}{}^\a (\cE
  dA)_{A_4A_5A_6]}{}^\b.
\ee
The Bianchi identities for the two-forms are
\be\label{bianchitwo}
D_{[A_1}(dA)_{A_2A_3A_4]}{}^\ta =0.
\ee

We can now directly verify our main claim, that the bosonic equations 
of motion of ${\rm IIB}$ supergravity when reduced to one (time)
dimension and the $E_{10}$ $\s$-model equations when truncated to
$|\ell|\leq 4$ coincide if one makes the following identifications 
between the $t$-dependent $\s$-model quantities up to $\ell =4$, and 
the IIB supergravity quantities evaluated at a fixed, but arbitrarily 
chosen spatial point ${\bf x} = {\bf x}_0$:
\be
n(t) &=& Ne^{-1} (t,{\bf x}_0),\\
{P}_{ab}^{(0)} (t) &=& e_{(a}{}^m \pt e_{mb)} (t,{\bf x}_0),\\
Q_{ab}^{(0)} (t)&=& e_{[a}{}^m \pt e_{mb]}(t,{\bf x}_0) ,\\
P^{(0)1} (t) &=& S^1_t (t,{\bf x}_0)= e^{\phi}\pt\chi (t,{\bf x}_0),\\
P^{(0)3} (t) &=& S^3_t(t,{\bf x}_0) = \pt\phi (t,{\bf x}_0) ,\\
Q^{(0)} (t)&=& R_t (t,{\bf x}_0) = e^\phi\pt\chi (t,{\bf x}_0),\\
P_{a_1a_2}^{(1)}{}^\a (t) &=& 
\cE_\a{}^\ta e_a{}^m e_b{}^n \pt A_{mn,\ta} (t,{\bf x}_0),
\ee
\be
P_{a_1\ldots a_4}^{(2)}(t) &=& e_{a_1}{}^{m_1}\cdots e_{a_4}{}^{m_4}
H_{tm_1\ldots m_4} (t,{\bf x}_0),\\
P_{a_1\ldots a_6}^{(3)}{}^\a (t) &=& \frac1{3!}n e \e_{a_1\ldots a_6
c_1\ldots c_3} {e_{c_1}}^{m_1} \cdots {e_{c_3}}^{m_3}
{\cE_\a}^\ta \partial_{m_1} A_{m_2 \dots m_3, \ta}  (t,{\bf x}_0) \\ 
P_{a_1\ldots a_7|a_8}^{(4)} (t) &=& \frac12 n e \e_{a_1\ldots a_7 bc}
{\tilde{\Omega}}_{bc|a_8} (t,{\bf x}_0) ,\\
P_{a_1\ldots a_8}^{(4)}{}^\hi (t) &=& n e \e_{a_1\ldots a_8b} S_b^\hi
 (t,{\bf x}_0),\\
P_{a_1\ldots a_8}^{(4)}{}^2 (t) &=& n e \e_{a_1\ldots a_8b} R_b^\hi
 (t,{\bf x}_0).
\ee
Here, $e=\det (e_m{}^a)$ and ${\tilde{\Omega}}_{ab|c}$ is the traceless 
part of the anholonomy,the trace part has been gauged to zero
\cite{Damour:2004zy}. Spatial derivatives of the lapse $N$  are
neglected in this approximation and hence, for example, out of the
three terms in the covariant derivate (\ref{covder}) involving the
spin connection only the second term survives. For the
$SL(2,\reals)/SO(2)$ coset we have adopted the parametrization
above. The $SL(2,\reals)$ symmetry plays an important part in this
identification: for instance, the formula for $P_{a_1a_2}^{(1)}{}^\a$
must contain the neunbein and the zweibein $\cE$ in precisely the
indicated form in order to be compatible with the symmetries.

With these identifications, eq.~(\ref{sigmaeomgrav}) coincides with the
Einstein equation (\ref{iibgrav}), if we recall from \cite{Damour:2004zy}
that $n\cD^{(0)}(n^{-1}P_{ab}^{(0)})$ can be directly identified with
the part of the spatial Ricci tensor containing only time derivatives,
$-N^{2}R_{ab}^{(0)}$,  cf. Eqn.~(4.66) of \cite{Damour:2004zy}. 
The contribution from $\ell=4$ on the r.h.s. of the $\ell=0$
equation of motion gives the leading terms of $R_{ab}$ in the 
first spatial derivatives of the vielbein, but there appears 
a mismatch involving the contribution from $P_{a_1\ldots a_8}^{(4)}{}^i$ 
to the Einstein equation, analogous to the one involving the
subleading terms from the curvature as in \cite{Damour:2004zy}. 
The coset eq.~(\ref{sigmaeomscalars}) is mapped to (\ref{iibcos}) 
in the reduction. The equations of motion (\ref{iibtwo}) and the 
Bianchi identities (\ref{bianchitwo}) are mapped to the $\s$-model 
equations (\ref{sigmaeoml1}) and (\ref{sigmaeoml3}), respectively. 
Equations (\ref{sigmaeoml4a}) and (\ref{sigmaeoml4b}) are related to 
the factorization of the vielbeine of the two cosets in the vicinity
of a space-like singularity~\cite{Damour:2004zy}.

Remarkably, the self-duality of the five-form field strength $H_5$ is 
built into $E_{10}$. The right hand side of (\ref{iibtwo}) is expanded to
\be
H_{bcdef}\e^{\a\b}(\cE dA)_{def}{}^\b
- 3 N^2 H_{bcdet}\e^{\a\b}(\cE dA)_{det}{}^\b.
\ee
The second term is precisely of the form of
the $[(\ell=2),(\ell=1)]$ term generated by $E_{10}$ in
(\ref{sigmaeoml1}) but the first term appears to be in conflict
with $E_{10}$. This puzzle is resolved by restricting $H_5$ to be
self-dual such that
\be
H_{bcdef}\e^{\a\b}(\cE dA)_{def}{}^\b
 =\frac1{4!}N^{-1}\e_{bcdef c_1\ldots
 c_4}H_{tc_1\ldots c_4} \e^{\a\b} (\cE dA)_{def}{}^\b,
\ee
and then is recognized as the $\big[(\ell=3),(\ell=2)\big]$ contribution to
(\ref{sigmaeoml1}). Likewise, matching eq.~(\ref{sigmaeoml2}) with
(\ref{iibfive}) requires the self-duality of the five-form field
strength and then can be read either as the Bianchi identity or the
equation of motion. Therefore, $E_{10}$ is related to ${\rm IIB}$
dynamically only for self-dual field strength of the five-form. 

\end{section}

\begin{section}{Conclusions}

As we have shown the geodesic action of the $E_{10}/K(E_{10})$
$\s$-model based on the standard bilinear form for $E_{10}$,
with manifest $SL(2,\reals)$ symmetry together with the representation
content up to $\ell\leq 4$ implies the self-duality constraint
of the five-form field strength of IIB supergravity. In the standard
derivation of the bosonic ${\rm IIB}$ equations of motion this
self-duality cannot be deduced from the bosonic symmetries, but becomes
necessary only when one adds fermions to make the theory locally
supersymmetric. In addition all the other fields of type ${\rm IIB}$
supergravity theory can be accommodated naturally, and their dynamics
is equivalent to those in the $\s$-model, at least up to the level
considered here.

In contradistinction to the non-hyperbolic algebra $E_{11}$, the
hyperbolic Kac Moody algebra $E_{10}$ does {\em not} allow for a
source of the D9-brane of IIB string theory in any natural
way~\cite{Kleinschmidt:2004dy,Kleinschmidt:2003mf}.
If one insists that the D9-brane is essential for string dualities, one
would therefore conclude that $E_{10}$ is `too small' \cite{Chaudhuri04}. 
The D9-brane couples to a ten-form potential, which in ten space-time 
dimensions has vanishing field strength and hence no dynamical degrees 
of freedom (a generalization of the IIB theory with a {\em doublet} of 
ten-form potentials has been given in \cite{Bergshoeff}). In eleven dimensions,
on the other hand, the equation of motion for such a rank-ten potential 
implies constancy of the associated field strength. This is a well-known 
mechanism for generating masses and a cosmological constant in (super-)gravity
\cite{ANT}. Yet, a cosmological term in eleven dimensions is inconsistent 
with 32 supersymmetries, and therefore such a modification of $D=11$
supergravity does not appear to exist \cite{BDHS}. Disregarding this 
fact for the moment, an appropriate field transforming in a singlet
representation of $SL(10)$ can be identified in an $A_9$ decomposition
of $E_{11}$ \cite{Kleinschmidt:2003mf,West}. In terms of a $D_{10}$
decomposition of $E_{11}$ analogous to \cite{Kleinschmidt:2004dy}, the
corresponding generator is part of an $SO(10,10)$ multiplet which
includes all RR potentials. To recover these representations from the
perspective  taken here, we have to decompose $E_{11}$ under
$A_9\times A_1$, with $SL(10)\subset SO(10,10)$. Indeed, one finds new
$SL(10)$ singlet representations at level $\ell=5$, which are absent
in $E_{10}$, and contain the D9-brane generator. However, the relevant
representation is an $SL(2,\reals)$ {\em quadruplet}. Therefore,
$E_{11}$ would predict the existence of {\em four} nine-brane objects,
transforming under $SL(2,\reals)$, whereas current superstring wisdom
seems to be compatible only with a {\em doublet} of such objects
\cite{Hull:1997kt,Meessen:1998qm,Bergshoeff}.  

Finally, we mention that it is believed that the continuous
$SL(2,\reals)$ symmetry of ${\rm IIB}$ supergravity is broken to
$SL(2,\ints)$ by quantum effects in the string theory. Similar effects
have recently been discussed
for the full $E_{10}(\reals)\rightarrow E_{10}(\ints)$ in
\cite{Brown:2004jb}. The inclusion of fermions along the lines of
\cite{Kleinschmidt:2004dy} seems straight-forward, with the local
$SO(9)$ appearing here being identified with the diagonal subgroup
of $SO(9,9)$.

\vspace*{0.5cm}
\noindent
{\bf Acknowledgements:} We are grateful to E. Bergshoeff and S. Chaudhuri 
for correspondence.

\end{section}

\end{document}